\documentclass[twocolumn,showpacs,preprintnumbers,amsmath,amssymb]{revtex4}

\usepackage{graphicx}
\usepackage{dcolumn}
\usepackage{bm}

\begin{document}

\title{Quantum Hall effects in fast rotating Fermi gases with anisotropic dipolar interaction}
\author{R.-Z. Qiu$^1$, Su-Peng Kou$^2$, Z.-X. Hu$^3$, Xin Wan$^{4}$, and S. Yi$^1$}
\affiliation{$^1$Key Laboratory of Frontiers in Theoretical Physics, Institute of
Theoretical Physics, Chinese Academy of Sciences, Beijing 100190, China}
\affiliation{$^2$Department of Physics, Beijing Normal University, Beijing 100875, China}
\affiliation{$^3$Department of Electrical Engineering, Princeton University, Princeton, New Jersey
08544, USA}
\affiliation{$^4$Zhejiang Institute of Modern Physics, Zhejiang University, Hangzhou 310027, China}
\date{\today }
\pacs{03.75.Ss, 73.43.-f, 73.43.Nq}

\begin{abstract}
We investigate fast rotating quasi-two-dimensional dipolar Fermi gases
in the quantum Hall regime. By tuning the direction of the dipole
moments with respect to the $z$-axis, the dipole-dipole interaction
becomes anisotropic in the $x$-$y$ plane. For a soft confining
potential we find that, as we tilt the angle of the dipole moments,
the system evolves from a $\nu = 1/3$ Laughlin state with dipoles
being polarized along the $z$ axis to a series of ground states
characterized by distinct mean total angular momentum, and finally to
an anisotropic integer quantum Hall state. During the transition from
the fractional regime to the integer regime, we find that the density
profile of the system exhibits crystal-like structures. We map out the
ground states as a function of the tilt angle and the confining
potential, revealing the competition of the isotropic confining
potential and both the isotropic and anisotropic components of the
dipole-dipole interaction.
\end{abstract}

\maketitle

\section{Introduction}

The quest for quantum computer with intrinsic fault tolerance spurs
recent interest in searching for exotic fractional quantum Hall (FQH)
states that support non-Abelian anyons~\cite{nayak08}. While it is
easy to write down a trial wave function with highly nontrivial
statistics, the realization of it in two-dimensional electron gases
(2DEGs) is not simple. Apart from the technical difficulties of sample
preparation and low operating temperature, the lack of effective
control on the interaction between particles is a major concern.  In a
realistic 2DEG, electrons interact via long-range Coulomb interaction,
which may be modified due to the presence of an adjacent gate in the
case of graphene or the effect of Landau level (LL) mixing, which
introduces effective three-body interaction (among others). In
particular, recent theoretical and experimental
studies~\cite{bishara09,xia10,wojs10,rezayi11} suggest that the
perturbative modification of the interparticle interaction can have
significant effects on the stability of FQH states. Therefore, the
question of how topological order evolves with interparticle
interaction remains an interesting question with growing experimental
capabilities of controlling microscopic parameters.

Ultracold atomic gases provide an ideal platform for simulating
quantum many-body systems~\cite{bloch}. The realizations of FQH states
in ultracold Fermi gases have been discussed in the presence of, for
example, a rapidly rotating trap~\cite{cooper,fetter} or a
laser-induced geometric gauge field~\cite{lin}. For identical
fermions, $s$-wave interactions vanish due to the Pauli exclusion
principle. Unless in the resonance regime, $p$-wave interactions in a
single-component Fermi gas are typical very small. Nevertheless,
significant interactions can still be introduced by using atoms or
molecules with strong dipole-dipole
interactions~\cite{lu,ye1,ye2}. The FQH effects in a two-dimensional
(2D) dipolar Fermi gas with isotropic dipole-dipole interaction have
been studied in Refs.~\cite{bara,Lewenstein}. The system has been
shown to undergo transitions from an integer quantum Hall (IQH) state to a
$\nu=1/3$ Laughlin state, and to a Wigner-crystal state by increasing
the rotation frequency. However, we are not aware of any work on the
FQH effects in the presence of anisotropic interaction, when the
dipoles are not oriented along the rotation axis.

In the present work, we study the FQH effects in a fast rotating
quasi-2D gas of polarized fermionic dipoles. By tilting the direction
of the dipole moments with respect to the rotation axis, we can tune
the dipole-dipole interaction to be anisotropic on the plane of
motion. Starting from a Laughlin state with isotropic dipolar
interaction, we investigate the ground state properties by varying the
tilt angle of the dipole moments. We find that for small tilt angle
the ground state can be approximately described by a FQH
state. However, as one further increases the tilt angle, the ground
state deviates from the FQH state significantly such that a
crystal-like pattern emerges in the density profile of the gas. The
ground state of the system eventually becomes an IQH state when dipole
moments are aligned in the 2D plane. For a soft confining potential,
the IQH state is noticeably anisotropic. We map out the phase diagram
in the parameter space spanned by the tilt angle and the strength of
the confining potential. The results can be explained by the
competition of the isotropic confining potential and both the
isotropic and anisotropic components of the dipole-dipole interaction.

This paper is organized as follows. In Sec.~\ref{formu}, we present
our model and calculation for relevant matrix elements of the model
Hamiltonian. Section~\ref{iso} briefly covers the FQH states with
isotropic dipolar interaction for later comparisons. In
Sec~\ref{aniso} we investigate the ground state structure in the
presence of anisotropic dipole-dipole interaction in a weak confining
potential. The full phase diagram is presented in Sec.~\ref{resu}. We
conclude our discussions in Sec.~\ref{conc}.

\section{Model}\label{formu}

We consider a system of $N$ spin polarized fermionic dipoles trapped in an axially symmetric potential
\[
U({\mathbf{r}})=\frac{1}{2}\mu(\omega ^{2}x^{2}+\omega
^{2}y^{2}+\omega _{z}^{2}z^{2}),
\]
where $\mu$ is the mass of the particle, $\omega $ and $\omega _{z}$ are the radial and axial trap frequencies, respectively. The trapping potential rotates rapidly around the $z$-axis with an angular frequency $\Omega$ ($<\omega $). We further assume that the dipole moments $d$ of all particles are polarized by an external orienting field which is at an angle $\theta $ about the $z$-axis. Since the $s$-wave collisional interaction vanishes for spin polarized fermions, particles only interact with each other via a dipole-dipole interaction. If the orienting field corotates with the trapping potential, the dipolar interaction becomes time-independent in the rotating frame, i.e.,
$$\mathcal{V}(\mathbf{r})=c_{d}V_{\theta }^{\mathrm{(3D)}}({\mathbf{r}}),$$ where $c_{d}=d^{2}/(4\pi \varepsilon _{0})$ or $\mu _{0}d^{2}/(4\pi )$ for, respectively, electric or magnetic dipoles, with $\varepsilon _{0}$ ($\mu _{0}$) being the vacuum permittivity (permeability). The spatial dependence of ${\cal V}({\mathbf r})$ can be described by
\[
V_{\theta }^{\mathrm{(3D)}}(\mathbf{r})=\frac{1}{r^{5}}\,\left[r^{2}-3(z\cos \theta +x\sin \theta )^{2}\right].
\]
Here, without loss of generality, we have assumed that the dipole moments are polarized in the $x$-$z$ plane of the rotating frame. We can tune the dipolar interaction by introducing a tilt angle $\theta $ such that $V_{\theta }^{\mathrm{(3D)}}({\mathbf{r}})$ is isotropic (anisotropic) on $x$-$y$plane for $\theta =0$ ($\theta \neq 0$). In the rotating frame, the Hamiltonian of the system becomes
\begin{eqnarray}
H_{\mathrm{3D}}=\sum_{i}\left[ \frac{{\mathbf{p}}_{i}^{2}}{2\mu}+U({\mathbf{r}}
_{i})-\Omega L_{i}^{z}\right] +c_{d}\sum_{i<j}V_{\theta }^{\mathrm{(3D)}}({
\mathbf{r}}_{i}-{\mathbf{r}}_{j}),\nonumber\\\label{hami1}
\end{eqnarray}
where $L^{z}=xp_{y}-yp_{x}$ is the $z$ component of the orbital angular momentum.

Under the condition $\omega _{z}\gg \omega $, the system can be regarded as quasi-2D. As a result, the motion of all particles along the $z$-axis is frozen to the ground state of the axial harmonic oscillator, with a wave function $\phi _{z}(z)=\pi^{-1/4}q^{-1/2}e^{-z^{2}/(2q^{2})}$, where $q=\sqrt{\hbar /(\mu\omega _{z})}$.
Integrating out the variable $z$ from Eq. (\ref{hami1}), we obtain the
Hamiltonian for the quasi-2D system as
\begin{eqnarray}
H_{\mathrm{2D}} &=&\sum_{i}\left[ \frac{({\mathbf{p}}_{i}-\mu\omega \hat{%
\mathbf{e}}_{z}\times {\boldsymbol{\rho }}_{i})^{2}}{2\mu}+\hbar (\omega
-\Omega )L_{i}^{z}\right]  \nonumber \\
&&+c_{d}\sum_{i<j}V_{\theta }^{\mathrm{(2D)}}({\boldsymbol{\rho }}_{i}-{%
\boldsymbol{\rho }}_{j}).  \label{hami2d}
\end{eqnarray}%
where ${\boldsymbol{\rho }}=(x,y)$, $\hat{\mathbf{e}}_{z}$ is the unit
vector along the $z$-axis and
\begin{eqnarray}
V_\theta^{\rm (2D)}({\boldsymbol{\rho}})=\frac{1}{(2\pi q^2)^{1/2}}\int dz e^{-z^2/(2q^2)} V_\theta^{(3D)}({\boldsymbol\rho},z).
\label{vdd2}
\end{eqnarray}
The first term on the righthand side of Eq.~(\ref{hami2d}) represents the single-particle Fock-Darwin Hamiltonian in the symmetric gauge~\cite{Fock,Darwin}, which can be solved exactly to yield eigenenergies~\cite{fetter}
\begin{eqnarray}
\hbar(\omega-\Omega)n_++\hbar(\omega+\Omega)n_-+\hbar\omega,
\end{eqnarray}
known as the Fock-Darwin levels, where the quantum numbers $n_+$ and $n_-$ are two non-negative integers. In the fast rotating limit $\Omega\rightarrow\omega$, the Fock-Darwin levels mimic the LLs with a level spacing $2\hbar\Omega$. Throughout this work, we assume that the interaction energy is much smaller than the LL spacing, such that particles only occupy the highly degenerate lowest Landau level (LLL).

To proceed further, it is convenient to introduce a set of dimensionless units: $\hbar$ for angular momentum, $\ell=\sqrt{\hbar/(2\mu\omega)}$ for length, and $c_{d}/\ell^{3}$ for energy. The wave function of the LLL can then be expressed as
\[
\psi _{m}(\rho,\varphi)=\frac{\rho ^{m}e^{im\varphi }e^{-\rho ^{2}/4}%
}{\sqrt{2\pi 2^{m}m!}}\quad (m\geq 0),
\]
describing a state with an angular momentum $m\hbar$. Within the LLL formalism, the Hamiltonian~(\ref{hami2d}) in the second quantization reads
\begin{equation}
H=\alpha L^{z}+\frac{1}{2}\sum_{m_{1}m_{2}m_{3}m_{4}}V_{1234}(\theta
)f_{m_{1}}^{\dag }f_{m_{2}}^{\dag }f_{m_{4}}f_{m_{3}},  \label{vx1234}
\end{equation}
where $f_{m}^{\dag }$ is the fermion creation operator that creates a particle in state $\psi _{m}$ and $L^{z}=\sum_{m}mf_{m}^{\dag }f_{m}$ the total angular momentum. The dimensionless quantity $\alpha =\hbar (\omega -\Omega )\ell ^{3}/c_{d}$ characterizes the relative strength of confining potential with respect to interaction. In the presence of anisotropy, the interaction matrix elements are
\begin{widetext}
\begin{eqnarray}
V_{1234}(\theta ) &=&\int d{\boldsymbol{\rho }}_{1}d{\boldsymbol{\rho }}%
_{2}\psi _{m_{1}}^{\ast }({\boldsymbol{\rho }}_{1})\psi _{m_{2}}^{\ast }({%
\boldsymbol{\rho }}_{2})V_{\theta }^{\mathrm{(2D)}}({\boldsymbol{\rho }}_{1}-{\boldsymbol{%
\rho }}_{2})\psi _{m_{3}}({\boldsymbol{\rho }}_{1})\psi _{m_{4}}({%
\boldsymbol{\rho }}_{2})\nonumber\\
&=&{\cal A}_{1234}\left[\frac{3\cos^2\theta-1}{2}
\left(\frac{8}{3}{\cal J}_{1234} -4{\cal
K}_{1234}\right)\delta_{m_1+m_2,m_3+m_4}
+\sin^2\theta{\cal K}_{1234}\delta_{m_1+m_2,m_3+m_4\pm2}\right],\label{v1234}
\end{eqnarray}
\end{widetext}
where
\begin{equation}
{\cal A}_{1234}=\frac{1}{2\sqrt{2\pi}q}\frac{i^{|m_3-m_1|-|m_4-m_2|}} {2^{(|m_3-m_1|+|m_4-m_2|)/2}}\sqrt{\frac{[m^<_{13}]![m^<_{24}]!}{[m^>_{13}]![m^>_{24}]!}}, \nonumber
\end{equation}
\begin{eqnarray}
{\cal J}_{1234}&=&\int d\rho \rho^{|m_3-m_1|+|m_4-m_2|+1}e^{-\rho^2}\nonumber\\
&&\!\!\!\!\times L_{m_{13}^<}^{|m_3-m_1|} \!\left(\frac{\rho^2}{2}\right)L_{m_{24}^<}^{|m_4-m_2|} \!\left(\frac{\rho^2}{2}\right),\nonumber
\end{eqnarray}
\begin{eqnarray}
{\cal K}_{1234}&=&q\sqrt{\frac{\pi}{2}}\int d\rho \rho^{|m_3-m_1|+|m_4-m_2|+2}e^{-\rho^2+q^2\rho^2/2}\nonumber\\
&&\!\!\!\!\times L_{m_{13}^<}^{|m_3-m_1|} \!\left(\frac{\rho^2}{2}\right)L_{m_{24}^<}^{|m_4-m_2|} \!\left(\frac{\rho^2}{2}\right){\rm erfc}\!\left(\frac{q\rho}{\sqrt{2}}\right),\nonumber
\end{eqnarray}
where $m_{ij}^{<}= \mathrm{min}(m_{i},m_{j})$ and $m_{ij}^{>}=\mathrm{max} (m_{i},m_{j})$. $L_{m}^{n}(\cdot)$ is the associated Laguerre polynomial and
$\mathrm{erfc}(\cdot)$ is the complementary error function. Clearly, for $\theta =0$ $V_{1234}$ is nonzero only when $m_{1}+m_{2}-m_{3}-m_{4}=0$, indicating that the total angular momentum $L_z$ is conserved in the isotropic case. However, when the interaction becomes anisotropic ($\theta \neq 0$), $V_{1234}$ are nonzero when $m_{1}+m_{2}-m_{3}-m_{4}=0$ or $\pm 2$. 

Hamiltonian (\ref{vx1234}) contains three parameters: the total number
of particles $N=\sum_{m}f_{m}^{\dag }f_{m}$, the relative strength of
the confining potential $\alpha$, and the tilt angle $\theta$ of the
dipole moment. In the following sections, we will explore the quantum
states of the system in the parameter space $(N,\theta ,\alpha )$. Our
main focus is on the parameter ranges of $N\leq 10$, $0\leq \theta\leq
\frac{\pi}{2}$, and $0.01\leq\alpha \leq 0.1$. Unless otherwise
stated, the value of $q$ is chosen to be $0.01\ell$.

\section{Quantum Hall states with isotropic dipolar interaction}\label{iso}

Let us first assume that the dipolar interaction is isotropic in the
$x$-$y$ plane, which corresponds to $\theta=0$ in the
Hamiltonian~(\ref{vx1234}). In this case, the total angular momentum
is conserved. Therefore, one may numerically diagonalize the
Hamiltonian~(\ref{vx1234}) in the subspace of a given total angular
momentum $M$ to obtain
\begin{eqnarray}
H\left|\Phi^{(N)}_{M,n}\right\rangle=E^{(N)}_{M,n}\left|\Phi^{(N)}_{M,n}\right\rangle,
\end{eqnarray}
where $E^{(N)}_{M,n}$ and $\left|\Phi^{(N)}_{M,n}\right\rangle$
are eigenenergies and eigenstates, respectively. The index $n$ labels
the state in the subspace of total angular momentum $M$ with
increasing eigenenergy, i.e., $n=0$ for the lowest energy state, $n=1$
for the first excited states, etc. We emphasize that we present a
numerically exact treatment of the dipolar interaction potential for a
quasi-2D system with finite wave-function spread along the
perpendicular direction, while the ideal 2D case has been studied
previously~\cite{Lewenstein}.

\begin{widetext}

\begin{table}[h]
\caption{Magic numbers in an isotropic $N=10$ system with various
  $q$s. The magic numbers obtained from the composite fermion theory
  are included for comparison.}
\begin{tabular}{l|ccccccccccccccccccccc}
\hline\hline
CF&45&&55&&63&&69&&77&&83&&90&&&97&103&111&117&125&135\\ 
\hline
$q=0.5$&45&52&&59&&66&69&73&77&80&&85&90&93&&97&103&111&117&125&135\\
\hline
$q=0.1$&45&52&&59&&66&&73&77&80&&85&90&93&95&&103&111&117&125&135\\
\hline
$q=0.01$&45&52&&59&&66&&73&77&80&&85&90&93&95&&103&111&117&125&135\\
\hline\hline
\end{tabular}
\label{table}
\end{table}

\end{widetext}

\begin{figure}
\centering
\includegraphics[width=3.4in]{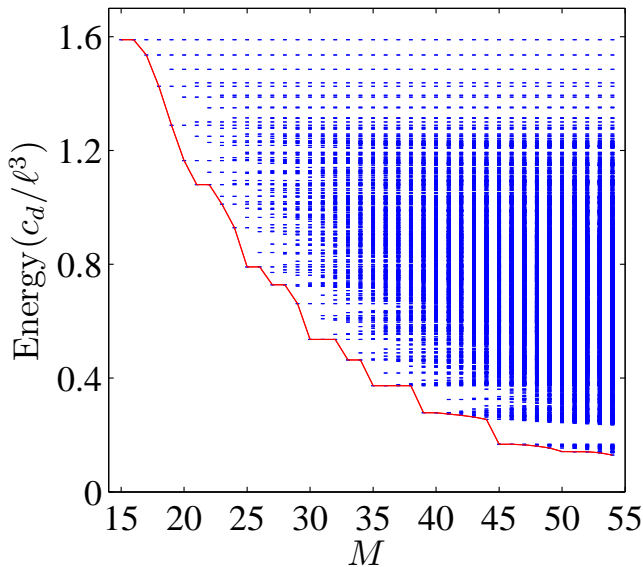}
\caption{(Color online) Energy spectrum of Hamiltonian~(\ref{vx1234})
  with $N=6$, $\alpha=0$, and $\theta=0$. The solid (red) line
  connects the lowest energy state in each total angular momentum
  subspace as a guide to the eyes. The smallest angular momentum state
  on the each shoulder of the line may become the global ground state
  of the system as $\alpha$ varies. } \label{speciso}
\end{figure}

In Fig.~\ref{speciso}, we plot the eigenenergy versus the total
angular momentum for $N=6$ and $\alpha=0$. As a guide to the eyes, we
have connected the lowest energy state in each total angular momentum
subspace by a piecewise straight line, on which a series of shoulders
are visible. The first state on the each shoulder, where a downward
cusp appears in the spectrum, represents a possible candidate for the
global ground state of the system with increasing $\alpha$. For a
given $\alpha\neq 0$, only one of these states is the global ground
state of the system; the corresponding total angular momentum is so
called a {\em magic number}. The properties of these states have been
studied extensively. Laughlin first noticed that the lowest-energy
state of the $M=3N(N-1)/2$ is closely related to the fractional
quantum Hall effect~\cite{Laughlinc}. To translate these numbers into
asymptotic filling factors in the thermodynamic limit, Girvin and Jach
proposed an explicit expression for these states in the finite
system~\cite{Girvinb},
\begin{eqnarray}
\label{eq:nu}
\nu=\frac{L_0}{M},
\end{eqnarray}
where $L_0=N(N-1)/2$. In the context of quantum dots, Jain and
Kawamura proposed an explanation of the magic numbers using the theory
of composite fermions~\cite{Jaina,Jainb}, which has been further
discussed in later works~\cite{Seki,Maksym,Landman}.  We list the
magic numbers of our quasi-2D model for $N=10$ in Table~\ref{table}
for several choices of $q$. For small $q$ up to 0.1, our results are
consistent with the results in 2D rotating Fermi gases with isotropic
dipolar interaction studied earlier by Osterloh {\it et
  al}~\cite{Lewenstein}. However, for larger $q$ we observe a few
discrepancies. Interestingly, the magic numbers 69 and 97, not showing
for small $q$, are consistent with the prediction of the composite
fermion theory.

\section{Quantum Hall states with anisotropic interaction}\label{aniso}
Now we turn to the study of the quantum states in systems with
anisotropic dipolar interaction. Since the total angular momentum
$L^z$ is no longer conserved, one has to numerically diagonalize the
Hamiltonian~(\ref{vx1234}) in much larger Hilbert spaces. In
practice, one may truncate the Hilbert space by introducing a cutoff,
$m_{\rm cut}$, to the angular momentum $m$, such that the
diagonalization procedures are carried out in the space constructed from
$\psi_m$s with $m\leq m_{\rm cut}$. We have chosen a sufficiently large
$m_{\rm cut}$ such that our results presented in this work are
not the choice of $m_{\rm cut}$. In numerical
diagonalization, we take advantages of the fact that dipole-dipole
interaction only couples angular momentum subspaces that differ by an
even angular momentum and diagonalize in the Hilbert space spanned by
odd and even angular momentum states separately.

For a given set of parameters $(N,\alpha,\theta)$, the ground state
wave function of the system is denoted by
$\left|\Psi^{(N)}(\alpha,\theta)\right\rangle$ with the ground state
energy $E^{(N)}(\alpha,\theta)$. Throughout this section, the strength
of the confining potential is fixed at $\alpha=0.01$, such that the
ground state is a $\nu=1/3$ Laughlin state for $\theta=0$. The
dependence of the results on $\alpha$ will be discussed in the next
section. Due to the large size of the Hilbert space involved in the
numerical diagonalization, We study systems of up to $N=8$ fermions.

\subsection{State transitions induced by varying $\theta$}\label{phatran}
\begin{figure}
\centering
\includegraphics[width=3.2in]{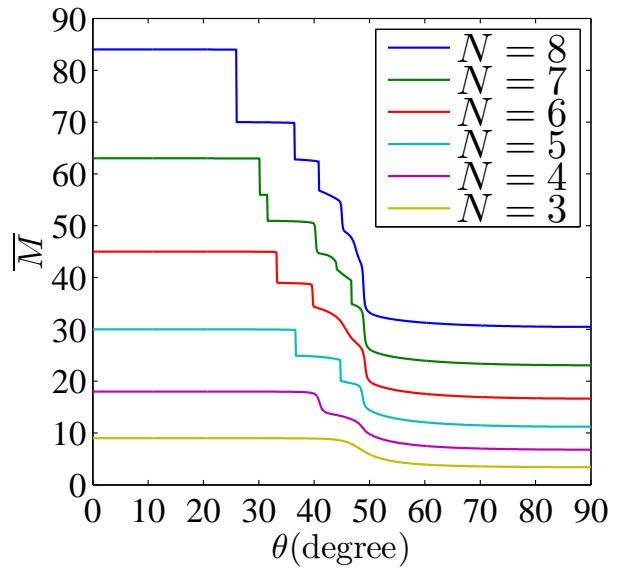}
\caption{(Color online) Mean total angular momentum $\overline M$
  versus the tilt angle $\protect\theta $ for $\alpha =0.01$ in
  systems with various sizes up to $N = 8$. For sufficiently large
  number of particles ($N \geq 5$), plateaus corresponding to distinct
  ground states develop on the curves. } \label{angular}
\end{figure}

Even though the total angular momentum $L^z$ is no longer a good
quantum number, its mean value,
\begin{eqnarray}
\overline M=\left\langle\Psi^{(N)}(\alpha,\theta)\right|L^z\left| \Psi^{(N)}(\alpha,\theta)\right\rangle,\nonumber
\end{eqnarray}
can be defined and we will see that it is sufficient to characterize
the ground state of a system. In Fig.~\ref{angular}, we plot the
$\theta$ dependence of $\overline M$ for systems with $N\leq 8$. We
find that $\overline M$ always decreases with $\theta$. As will be
shown, this monotonically decreasing behavior of $\overline M(\theta)$
is because the dipolar interaction becomes less repulsive as $\theta$
is increased, which reduces the size of the
system~\cite{Laughlinb}. Interestingly, for sufficiently large number
of particles ($N\geq 5$), plateaus develop on the $\overline
M(\theta)$ curves. Roughly speaking, for $\theta \lesssim 40^{\circ}$
the mean total angular momentum decreases abruptly from one plateau to
another as the tilt angle is increased, signaling sharp transitions
between ground states with distinct properties at various $\theta$.
For a concrete example, we examine in detail the transitions for the
system of $N=6$. The sudden drops of $\overline M$ are observed as
\begin{eqnarray}
\overline M:\quad 45\; {\stackrel{33.3^\circ}{\longrightarrow}} \;39.0 \;{\stackrel{39.9^\circ}{\longrightarrow}} \;34.89,\label{tran6}
\end{eqnarray}
where the numbers above the two arrows denote the critical tilt
angles. From the analysis presented in Sec.~\ref{iso}, the first and
second plateaus clearly correspond to the fractional quantum Hall
states with filling factors $\nu=1/3$ and $5/13$, respectively, in the
notation define in Eq.~(\ref{eq:nu}) by Girvin and
Jach~\cite{Girvinb}.  The third plateau has an mean total angular
momentum of 34.89, which is 0.3\% smaller than $35$, indicating that
it mainly contains the FQH state with filling factor $\nu=3/7$. As one
further increases $\theta$, $\overline M$ decreases smoothly toward an
asymptotic value of $17$ at $\theta=90^\circ$. As shown in
Fig.~\ref{angular}, similar features also appear in the systems with
$N=5$, $7$ and $8$. For $N=3$ and $4$, however, $\overline M$ always
varies smoothly, consistent with the expectation that quantum phase
transitions happen only in thermodynamically large systems; the
absence of the sharp steps for $N<5$ is the normal finite-size
artifact.

\begin{figure}
\centering
\includegraphics[width=3.in]{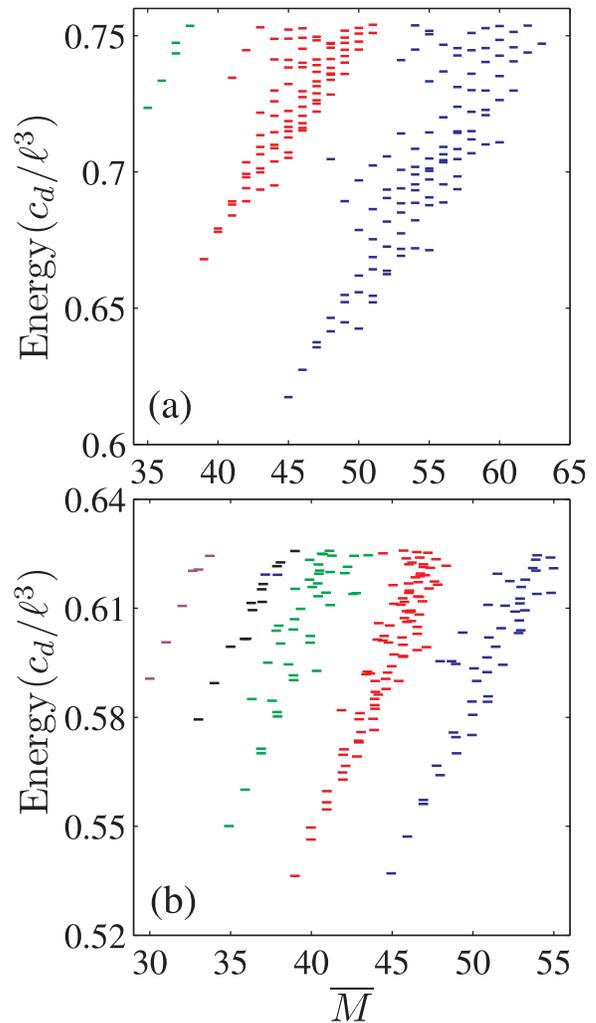}
\caption{(Color online) Low-lying energy spectrum for an $N=6$ system
  with $\alpha=0.01$ and a tilt angle of (a) $\theta =0^{\circ}$ and
  (b) $33.3^{\circ }$. Distinct colors are used to specify different
  energy-level clusters, which can be regarded as chiral excitations
  of the corresponding lowest-energy state.} \label{spectrum}
\end{figure}

To gain insight into those transitions, we examine the energy spectrum
of the system. In Fig.~\ref{spectrum}, we plot the low-lying energy
levels versus the mean total angular momentum for $N=6$ for the
isotropic and the anisotropic cases. In the isotropic case
($\theta=0$), the mean total angular momentum $\overline{M}$ is simply
the total angular momentum $L^Z$, which is a good quantum number.  In
both cases, energy levels group into clusters which are plotted with
different colors. For convenience, we shall refer to each energy
cluster using the angular momentum of the lowest energy level in the
cluster. Even though the boundary between two adjacent clusters are
not well defined for high-energy states, the lowest ones are clearly
well separated. In Fig.~\ref{spectrum}(a), we choose the range in
which three energy-level clusters are visible. Among them, the lowest
energy state in cluster-$45$ represents the ground state of the system
for $\theta=0$. This is the 6-particle Laughlin state as discussed in
Sec.~\ref{iso}. Apart from its mean total angular momentum, we can
also observe the feature of a Laughlin state also in the low-energy
spectrum. First of all, the low-lying excitations are chiral,
appearing only on the side of $\overline{M} > 45$. In particular, at
roughly $\overline{M} = 46$ and $47$, we find one and two states,
respectively. They can be interpreted as the edge excitations of the
ground state droplet, with their wave functions approximated by the
ground state wave function multiplied by a symmetric polynomial of the
corresponding degree. Near $\overline{M} = 48$, we expect three
low-lying states but only find two; however, there is another level
well above (near 0.7), presumably due to the influence of the cutoff
in the momentum space. The series of numbers of the low-lying states
is consistent with the chiral Luttinger liquid theory and signifies
the topological order of the corresponding ground state~\cite{7}.

As one increases the tilt angle $\theta$, all energy-level clusters
move downward to lower energies because the dipolar interaction
becomes less repulsive. Nevertheless, the counting of the low-lying
excitations remains robust, suggesting the topological order is not
destroyed by small anisotropy, as shown in Fig.~\ref{spectrum}(b). In
addition, the clusters with lower angular momentum move faster than
those with higher angular momentum, hence, for example, the lowest
energy state in cluster-$39$ becomes the ground state of the system at
$\theta=33.3^\circ$. As one further increases $\theta$, the lowest
energy state in cluster-35 will become the ground state of the
system. They have different low-energy excitation structure from that
of the $\overline{M} = 45$ ground state. In the case of the
cluster-$39$, in particular, there are two chiral excitations at
$\Delta \overline{M} \approx 1$, clearly different from the Laughlin
case.

\begin{figure}
\centering
\includegraphics[width=3.2in]{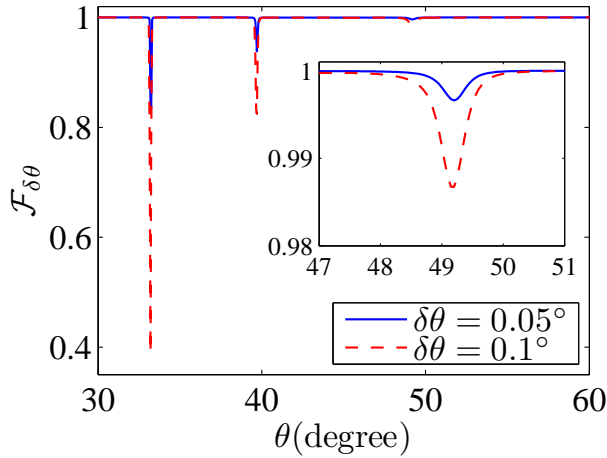}
\caption{(Color online) Fidelity ${\cal F}_{\delta\theta}$, as defined
  in Eq.~(\ref{eq:fidelity}), as a function of $\theta$ for $N=6$ and
  $\alpha=0.01$. The inset magnifies the fidelity in the vicinity of
  the small dip around $\theta = 49.3^{\circ}$.} \label{fidelity}
\end{figure}

The phase transitions induce by tuning anisotropy can be further
confirmed by calculating the fidelity of the ground state wave
function~\cite{quan,gusj},
\begin{equation}
\label{eq:fidelity}
{\cal F}_{\delta\theta}=\left|\left\langle\Psi^{(N)}(\alpha,\theta)
\right.\left|\Psi^{(N)}(\alpha,\theta+\delta\theta)\right\rangle\right|,
\end{equation}
where $\delta\theta$ is a small quantity. The fidelity $\cal F$
measures the similarity between two adjacent states in the parameter
space~\cite{gusj}. In the bulk of a single quantum phase, two states
close in the parameter space have wave functions that are only
perturbatively different, hence ${\cal F}_{\delta\theta}$ is close to
unity for sufficiently small $\delta\theta$. Near the phase boundary,
two states that are close in the parameter space can have very
different structure in their wave functions, therefore the fidelity
can drop sharply in the quantum critical region, signaling a quantum
phase transition in the thermodynamic limit. In Fig.~\ref{fidelity},
we plot the $\theta$ dependence of ${\cal F}_{\delta\theta}$ for a
systems with $N=6$ particles. Two transition points can be clearly
identified and the critical $\theta$ values are consistent with those
obtained in Fig.~\ref{angular}. A closer look at the ${\cal
  F}_{\delta\theta}(\theta)$ (inset of Fig.~\ref{fidelity}) further
reveals that there exists a third dip at $\theta\simeq
49.3^\circ$. The comparison of the fidelity ${\cal F}_{\delta\theta}$
for two different $\delta\theta$s indicates that the similarity of the
ground states decreases with the increasing distance between the
parameter $\delta\theta$ near the dip.

\subsection{Structure of the ground state wave function with anisotropic interaction}
\begin{figure}
\centering
\includegraphics[width=3.in]{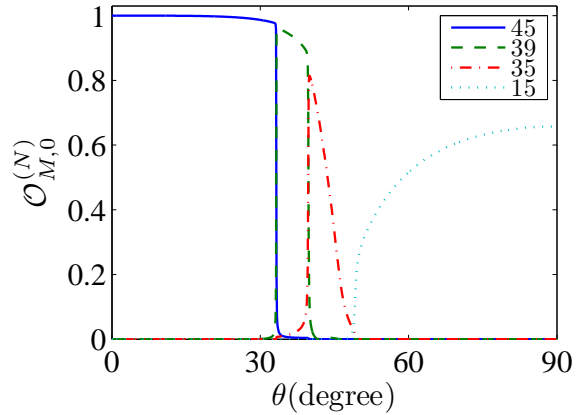}
\caption{(Color online) Overlap integral ${\cal O}^{(6)}_{M,0}$
  [Eq.~(\ref{eq:overlap})] versus $\theta$ for $M=45$ ($\nu = 1/3$
  Laughlin state), 39, 35, and 15 (IQH state) in the $N = 6$ system
  with $\alpha = 0.01$. } \label{overlap}
\end{figure}

To reveal the structure of the ground state wave function
$\left|\Psi^{(N)}(\alpha,\theta)\right\rangle$, let us calculate the
overlap integral
\begin{eqnarray}
\label{eq:overlap}
{\cal
  O}_{M,n}^{(N)}(\alpha,\theta)=\left|\left\langle\Phi^{(N)}_{M,n}\right|
\left.\Psi^{(N)}(\alpha,\theta)\right\rangle\right|
\end{eqnarray}
between the isotropic and anisotropic ground states. Again, we present
the data of the system with $N=6$ particles. Figure~\ref{overlap}
shows the dependence of ${\cal O}_{M,0}^{(N)}$ on anisotropy for $M =
45$, 39, 35, and $15$. As can be seen, the $\theta$-axis is roughly
divided into four regions, the boundaries of which coincide with the
three dips of the fidelity in Fig~\ref{fidelity}.

In the first region $\theta<33.2^\circ$, the overlap ${\cal
  O}_{45,0}^{(6)}$ is greater than $0.974$, which indicates that the
dominate contribution to the ground state wave function comes from the
Laughlin state $\left|\Phi^{(6)}_{45,0}\right\rangle$. Nevertheless,
close to the right boundary of this region, several states in $M=47$
and $43$ manifolds are mixed into the ground state wave function, such
that ${\cal O}^{(6)}_{45,0}$ drops notably. The ground state wave
function in region $33.3^\circ<\theta<39.6^\circ$ mainly contains the
states from $M=43, 41, 39$, and $37$ manifolds. Particularly, the
state $\left|\Phi^{(6)}_{39,0}\right\rangle$ provides the largest
contribution to the ground state with $0.858 \leq {\cal
  O}^{(6)}_{39,0} \leq 0.96$.

\begin{figure}
\centering
\includegraphics[width=3.2in]{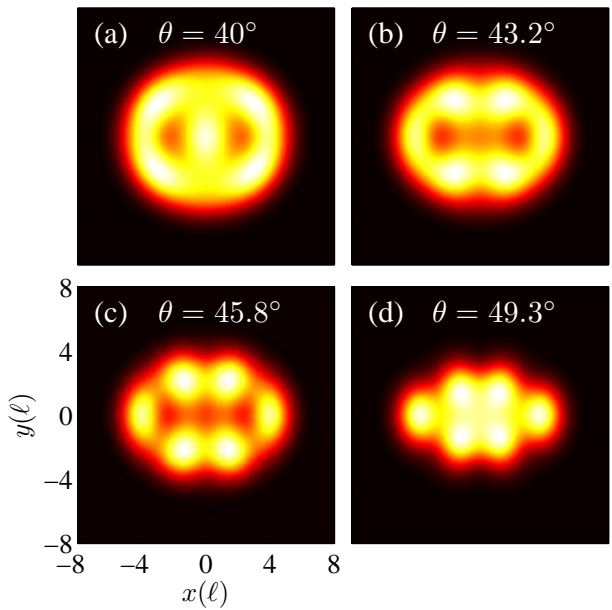}
\caption{(Color online) Density profiles $\varrho(x,y)$ for various
  $\theta$s in the $N=6$ system with $\alpha=0.01$. The brightest
  points indicate the local maxima in the density of the
  system. } \label{crys1}
\end{figure}

In the region $39.8^\circ<\theta<49.3^\circ$, the situation is much
more complicated compared to those in the first two regions. At the
left boundary, ${\cal O}_{35,0}^{(6)}$ is as high as $0.814$, but it
quickly drops to close to zero as the right boundary is approached. In
fact, many states from $M=21$ to $37$ in the isotropic case contribute
collectively to the ground state wave function. As a result,
complicated structures are developed in the density profile of the
system. In Fig.~\ref{crys1}, we present four typical patterns in the
density profiles
\begin{eqnarray}
\varrho(x,y)&=&\sum_{mm'}\left\langle\Psi^{(N)}(\alpha,\theta)\right|f_m^\dag f_{m'} \left|\Psi^{(N)}(\alpha,\theta)\right\rangle\nonumber\\
&&\quad\times\psi_m^*(x,y)\psi_{m'}(x,y)
\end{eqnarray}
of the system in this region. For $\theta=40^\circ$
[Fig.~\ref{crys1}(a)], besides the commonly seen ring-shaped density,
a vertical ridge appears along the $y$ axis. Five local density maxima
can be identified in the figure: 4 of them on the ring and the other
at the center of the trap. Figure~\ref{crys1}(b) shows the result for
$\theta=43.2^\circ$, on which 6 local density maxima appear on the
ring structure. Compared to the case at $\theta=40^\circ$, the density
profile is clearly stretched along the $x$ axis, which also represents
a generic trend for the density profile as $\theta$ is increased. The
reason behind this is because the interaction energy is lowered by
stretching the gas along the direction of dipole moment. As $\theta$
is further increased to $45.8^\circ$ [Fig.~\ref{crys1}(c)], the
structure with 6 local density maxima becomes more prominent, such
that each of them almost becomes an isolated island, as in a crystal
structure. In Fig.~\ref{crys1}(d), the 6 density islands start to
merge. We remark that these crystal-like structures appear as a result
of the interference between the states
$\left|\Phi_{M,n}^{(N)}\right\rangle$; the system is struggling to
maintain a balance between a large set of competing states. We note
that our calculations are based on a finite size system, this region
of competition may shrink in the thermodynamic limit, since more
plateaus are developing as system size increases as shown in
Fig.~\ref{angular}.

\begin{table}[h]
\caption{Largest 5 overlaps of the anisotropic state
  $\left|\Psi^{(6)}(\alpha,90^\circ)\right\rangle$ with the
  corresponding eigenstates in the isotropic case ($\theta = 0$). The
  pair of ($M$, $n$) labels the $n$th lowest eigenstate in the total
  angular momentum $M$ subspace. The system has $n = 6$ particles and
  $\alpha = 0.01$. }
\begin{tabular}{c c c c c c}
\hline\hline
$(M,n)$\hspace{0.2cm}&\hspace{0.2cm}$(15,0)$\hspace{0.2cm}&\hspace{0.2cm}$(17,0)$\hspace{0.2cm}&\hspace{0.2cm}$(19,2)$\hspace{0.2cm}&\hspace{0.2cm}$(21,7)$\hspace{0.2cm}&\hspace{0.2cm}$(23,14)$\\ 
\hline
${\cal O}^{(N)}_{M,n}$&0.658&0.608&0.387&0.196&0.084\\
\hline\hline
\end{tabular}
\label{table2}
\end{table}

\begin{figure}
\centering
\includegraphics[width=3.2in]{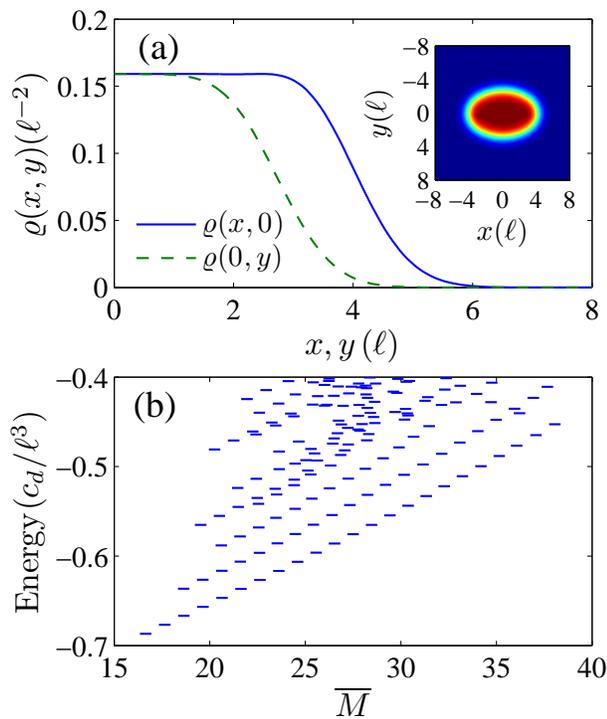}
\caption{(Color online) Characterization of the anisotropic IQH state
  at $\theta=90^\circ$ for $N=6$ and $\alpha=0.01$. (a) The density
  profiles along the $x$- and $y$-axes highlight the anisotropic
  nature of the state. Near the center, the LLL is completely
  filled. The inset shows the ellipsoidal density profile on the
  $x$-$y$ plane. (b) The low-lying energy levels can be counted as
  $1$, 1, 2, 3, 5, $\ldots$ for $\Delta \overline{M} = 0$, 1, 2, 3, 4,
  $\ldots$, respectively, which is consistent with the edge theory of
  an IQH state. } \label{theta90}
\end{figure}

In the last region $49.3^\circ<\theta\leq 90^\circ$, the most
important contributions to the ground state wave function are provided
by the states $\left|\Phi_{15,0}^{(6)}\right\rangle$,
$\left|\Phi^{(6)}_{17,0}\right\rangle$,
$\left|\Phi^{(6)}_{19,2}\right\rangle$, and
$\left|\Phi^{(6)}_{21,7}\right\rangle$. The probability of finding the
system in either of the above 4 states is larger than $0.92$
throughout this region. To understand the properties of the ground
state in this region, we first consider the specific state at
$\theta=90^\circ$. Table~\ref{table2} lists the largest values of
${\cal O}_{M,n}^{(6)}(0.01,90^\circ)$. Clearly, the main components of
$\left|\Psi^{(6)}(0.01,90^{\circ})\right\rangle$ are the $\nu=1$
IQH state $\left|\Phi_{15,0}^{(6)}\right\rangle$ and
its edge states, which suggests that, at $\theta=90^\circ$, the ground
state is an IQH state. To confirm this, we plot the
density distribution $\varrho(x,y)$ of this state in
Fig. \ref{theta90}(a). As can be seen, the surface density for the
elliptical plateau is exactly $1/2\pi \ell ^{2}$, which is identical
to that of a $\nu=1$ IQH state. Since the filling
factor can be defined as $\nu=2\pi\ell^2n_f$ with $n_f$ being the
fermionic surface density~\cite{Lewenstein}, the filling factor for
the $\theta=90^\circ$ state is thus $1$. The homogeneous elliptical
plateau of the density profile implies that our conclusion can be
generalized to the thermodynamic limit. We plot the energy spectrum of
the system in Fig.~\ref{theta90}(b). It is well-known that the
topological properties of the $\nu=1$ IQH state is
labeled by the number of edge states, i.e., $1,1,2,3,5,\ldots$ for
$\Delta L^z=0,1,2,3,4,\ldots$, which is exactly the case shown in
Fig.~\ref{theta90}(b), although in this case we have to replace
$\Delta L^z$ by $\Delta \overline{M}$. All the above evidences mount
to the fact that the $\theta=90^\circ$ state is a $\nu=1$ integer
quantum Hall state. Since there is no phase transition observed in
this region from various criteria, we conclude that the ground state
for $\theta>49.3^\circ$ can be characterized as an anisotropic integer
quantum Hall state.

\subsection{Understanding the anisotropic dipolar interaction}
\label{iso_aniso}
To understand how anisotropy in the dipolar interaction leads to the
emergence of the anisotropic IQH state, we decompose
the dipolar interaction into isotropic and anisotropic (on $x$-$y$
plane) components as
\begin{eqnarray}
V_\theta^{(3D)}&=&V_{\theta,\rm iso}^{(3D)}+V_{\theta,\rm{ani}}^{(3D)}\nonumber\\
&=&\eta_{\rm iso}(\theta)\frac{r^2-3z^2}{r^5}+\eta_{\rm ani}(\theta)\frac{
y^{2}-x^{2}}{r^{5}},\label{decomp}
\end{eqnarray}
where $\eta_{\rm iso}(\theta)=(3\cos^2\theta-1)/2$ and $\eta_{\rm
  ani}(\theta)=3\sin^2\theta/2$ represent the strengths of the
isotropic and anisotropic components, respectively. One should note
that, to obtain Eq.~(\ref{decomp}), we have neglected the linear terms
in $z$ as they vanish after we integrate out the variable $z$ to
obtain the 2D interaction potential. The properties of the system can
be seen as determined by the competition between $V_{\theta,\rm
  iso}^{(3D)}$ and $V_{\theta,\rm{ani}}^{(3D)}$. Since $\eta_{\rm
  iso}$ ($\eta_{\rm ani}$) is a decreasing (increasing) function of
the tilt angle, varying $\theta$ will change the relative strength of
the isotropic and anisotropic components, which can give rise to
different quantum phases.

\begin{figure}[tbp]
\centering
\includegraphics[width=3.2in]{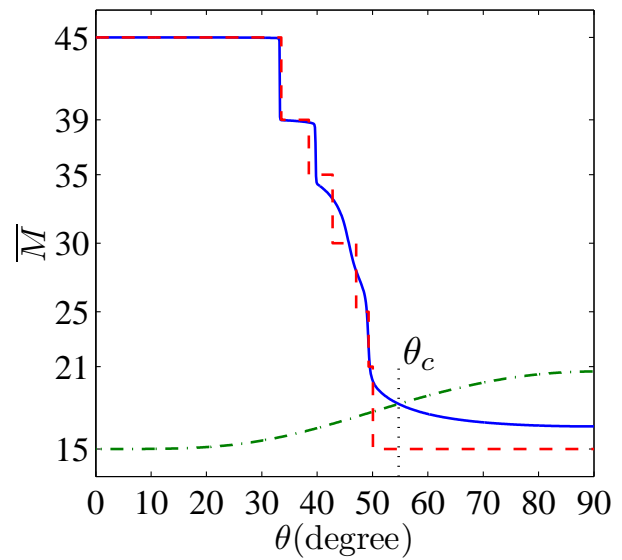}
\caption{(Color online) Mean total angular momentum as a function of
  $\theta$ for $N=6$ and $\alpha=0.01$ in systems of various
  interactions. The solid, dashed, and dash-dotted lines correspond
  to, respectively, the real system with dipolar interaction, FS-I
  with the isotropic component only, and FS-II with the anisotropic
  component only. The vertical dotted line indicates the position of
  angle $\theta_c$, where the isotropic component becomes attractive.}
\label{fake}
\end{figure}

After introducing the decomposition Eq.~(\ref{decomp}), we explore the
contributions of the isotropic and anisotropic components of the
dipolar interaction separately. To this end, we consider two
fictitious systems, FS-I and FS-II, in which the full dipolar
interaction is replaced, respectively, by $V_{\theta,\rm iso}^{(3D)}$
and $V_{\theta,\rm ani}^{(3D)}$. The corresponding Hamiltonians take
the same form as Eq.~(\ref{vx1234}) except for that the interaction
matrix elements are replaced by those calculated using $V_{\theta,\rm
  iso}^{(3D)}$ or $V_{\theta,\rm ani}^{(3D)}$.

In FS-I, the strength of the isotropic interaction, $\eta_{\rm
  iso}(\theta)$ decreases with $\theta$. Therefore, increasing the
tilt angle is effectively equivalent to increasing the strength of the
confining potential $\alpha$ for the full Hamiltonian with
$\theta=0$. We plot, in Fig.~\ref{fake}, the $\theta$ dependence of
the ground state angular momentum for system-I with $N=6$ and
$\alpha=0.01$. As $\theta$ is varied, FS-I experiences the same
transitions as those studied in Sec.~\ref{iso}. For the first two
transitions, the critical values of $\theta$ roughly agree with those
obtained using full dipolar interaction potential. In particular,
$V_{\theta,{\rm iso}}^{(3D)}$ vanishes at angle $\theta_c=54.74^\circ$
and becomes attractive in the $x$-$y$ plane for
$\theta>\theta_c$. Consequently, the ground state becomes a $\nu=1$
IQH state for $\theta\geq\theta_c$. The reason that the transition to
the IQH state occurs at a tilt angle smaller than $\theta_c$ is due to
the finite $\alpha$ used in Fig.~\ref{fake}.

In FS-II, we note that the system must be in the $\nu=1$ integer
quantum state at $\theta=0$ where $V_{\theta,\rm ani}^{(3D)}$
vanishes. As one increases $\theta$, the mean total angular momentum
$\overline M$ increases smoothly from $15$ to roughly $21$
(Fig.~\ref{fake}), which suggests that the system always stays in the
IQH state independent of the tilt angle, although the quantum Hall
droplet is stretched gradually along the $x$ axis. Further study on
the ground state of FS-II can be carried out as those have been done
in the previous subsection.

\begin{figure}[tbp]
\centering
\includegraphics[width=3.2in]{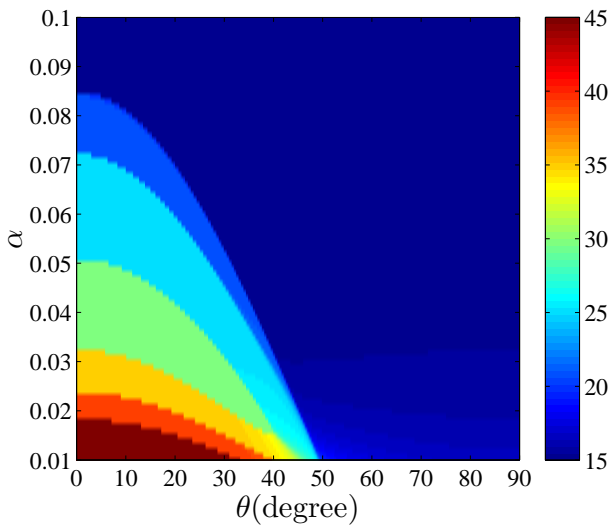}
\caption{(Color online) Mean total angular momentum $\overline
  M(\theta,\alpha)$ of the ground state for a system with $N=6$
  particles. The lower left corner is the $\nu = 1/3$ Laughlin phase,
  while the large (blue) region containing the upper right corner is
  the IQH phase, which crosses over to an anisotropic IQH phase at the
  lower right corner. }
\label{phase}
\end{figure}

From the above analysis, it becomes clear that the series of ground
state transitions induced by varying the tilt angle are mainly caused
by the isotropic component of the dipolar interaction. The anisotropic
component, on the other hand, changes the fine details of the ground
state, as it mixes in states with different total angular momentum to
the otherwise isotropic ground state, as is clearly exemplified in the
anisotropic integer quantum Hall state.

\section{Global phase diagram}\label{resu}

In Secs.~\ref{iso} and \ref{aniso}, we have shown the ground state
transitions induced by varying either the confining strength $\alpha$
or the tilt angle $\theta$. We present the complete phase diagram in
Fig.~\ref{phase}, in which we plot the mean total angular momentum
$\overline M$ as a function of $\theta$ and $\alpha$ for $N=6$. The
phase diagram is separated into regions with different value of
$\overline M$, which are well defined along the $\alpha$-axis, where
the interaction is isotropic, as discussed in Sec.~\ref{aniso}.
Within our numerical capabilities, we find the basic structure of
Fig.~\ref{phase} remains unchanged as system size $N$ varies.

We find that the Laughlin state with $\nu = 1/3$ is robust for weak
confinement and not too large tilt angle ($< 30^{\circ}$), which
assures that unintentionally introduced anisotropy in interparticle
interaction is not important. On the other hand, FQH cannot survive in
the large anisotropy of the dipolar interaction, when the isotropic
component of the interaction becomes soft, as analyzed in
Sec.~\ref{iso_aniso}.

When the confining strength becomes stronger, the mean total angular
momentum $\overline M$ becomes smaller, indicating the system of
particles becomes denser and denser. The evolution of $\overline M$ is
not smooth, but goes through a series of magic numbers, which can be
interpreted by the corresponding filling factors. In the large
confinement limit, the system develops into a maximum density droplet
with $\nu = 1$. The state crosses over to an anisotropic maximum
density droplet for large tile angles (at small confinement since the
isotropic component of the interaction changes from repulsive to
attractive interaction), as revealed in Fig.~\ref{theta90}, which
reflects the competition of the isotropic confining potential and both
the isotropic and anisotropic components of the interaction.

\section{Conclusion}\label{conc}

To summarize, we investigated the quantum Hall effects in the LLL of a
fast rotating quasi-2D Fermi gas with anisotropic dipolar interaction
through exact numerical diagonalization. With the tilt angle of the
dipole moment $\theta$, we introduced a new control knob to explore
the FQH effect in cold atomic systems. We studied in details the phase
diagram of a finite-size system and concluded that phase
transition is expected as the tilt angle $\theta$ varies in the
thermodynamic limit. When the tilt angle is small, the ground state of
the system can be described by a FQH state, whose filling fraction
depends on the strength of the confinement and hence the average
density. At large tilt angle, the system eventually becomes an
anisotropic $\nu=1$ IQH state. However, for
intermediate $\theta$ value, we find that crystal-like order develops
in the density profile of the system, suggesting the competition
between various parameters and phases. By decomposing the dipolar
interaction into isotropic and anisotropic components, we provide a
simple explanation to quantitatively understand the phase
transitions induced by anisotropy. The various competing orders and
ground states are summarized in a complete phase diagram in the
parameter space spanned by the tilt angle and the confinement. 

In the presence of anisotropy in the interparticle interaction, we
lose the rotational symmetry when treating the system in a disk
geometry, hence the total angular momentum is not a good quantum
number any more. Nevertheless, we demonstrated that one can still
compute the expectation value of the total angular momentum operator
for eigenstates and use it to characterize the various ground states
emerged as the results of competitions between confinement and
anisotropy, as well as between the isotropic and anisotropic
components of the interparticle interaction. Particularly interesting
is that the resilient features of topological order, such as the
presence of hierachical FQH ground states and the low-energy
excitations pertaining to the density deformation along the edge of an
incompressible quantum Hall droplet, remain robust in the presence of
anisotropy. While we showed that the calculation of the fidelity of
the ground state can be used as a probe to detect phase transitions
between states with different topological order, we believe the mean
angular momentum treatment can be readily generalized to the
calculation of, e.g., entanglement spectrum~\cite{li08}, which also
facilitates the detection of topological order.

In this paper we demonstrated that an incompressible FQH state with a
large excitation gap can survive a fairly large amount of anisotropy.
This implies that the Laughlin state, the exact ground state produced
by the hard-core potential, is stable against the anisotropic
perturbation. The finding is not unexpected given the remarkable
stability of the Laughlin states in the presence of long-range
interaction, finite thickness of the two-dimensional electron gas, and
disorder~\cite{wan05}. It is, however, intriguing to ask the effects
of anisotropy on more exotic non-Abelian quantum Hall states. A more
challenging question would be whether it is possible, by tuning the
anisotropy, to enhance a certain FQH state, hopefully with exotic
statistics. The present paper paves a path toward these questions.

\section*{ACKNOWLEDGMENTS}
This work was supported by the NSFC (Grant Nos. 11025421, 10935010,
10874017 and 10974209), the ``Bairen" program of the Chinese Academy
of Sciences, the DOE grant No. DE-SC0002140 (Z.X.H.) and the 973
Program under Project Nos. 2009CB929100 (X.W.) and 2011CB921803 (S.P.K).

\end{document}